\documentclass[preprint,showpacs,preprintnumbers,amsmath,amssymb]{revtex4}
\usepackage{graphicx}
\begin{document}
\title{Properties of vector mesons in four pseudoscalars electroproduction}

\author{G. Toledo S\'anchez}

\affiliation{ Instituto de F\'{\i}sica, Universidad Nacional Aut\'onoma de M\'exico. M\'exico D. F. C.P. 04510}
\date{\today}
\begin{abstract}
Vector mesons and the $W$ gauge boson share some common features, such as the spin and a very short lifetime. The electromagnetic properties of the $W$ are linked to its gauge nature and thus a big effort has been devoted to measure them, while accounting for its instability. 
In this work, we elaborate on how these ideas can be translated to study vector mesons. We focus in the unstable property of such states and the restrictions from electromagnetic gauge invariance, describing the differences and similarities. Then, we describe the four pseudoscalars electroproduction as an analog to the $e^+e^- \to 4 fermions$ process, used to study the electromagnetic properties of the $W$ boson. We point out that the current experimental capabilities are reaching the possibility to measure the magnetic dipole moment of light vector mesons.
 \end{abstract}

\maketitle

\section{Introduction}

The trilinear gauge boson interactions $WW\gamma$ and $WWZ$ come from the kinetic term of the SU(2) gauge field, upon rewriting it in terms of the mass eigenstates. Thus, it probes the non-abelian nature of the standard model. The corresponding $\gamma(q,\lambda)\to W^{-}(k_{-},\nu)W^{+}(k_{+},\mu)$ vertex, takes the form:
\begin{equation}\label{gammaWW}
  ie\left[g^{\mu\nu}\left(k_{-}-k_{+}\right)^{\lambda}+g^{\nu\lambda}\left(-q-k_{-}\right)^{\mu}+g^{\lambda\mu}\left(q+k_{+}\right)^{\nu}\right],
\end{equation}
which can be set, using momentum conservation and taking the $W$ particles on-shell as:
\begin{equation}\label{funcWWgamma_mod_std}
\Gamma_{0}^{\lambda\mu\nu}=g^{\mu\nu}\left(k_{-}-k_{+}\right)^{\lambda}+2\left(q^{\nu}g^{\lambda\mu}-q^{\mu}g^{\nu\lambda}\right),
\end{equation}
where we omit the $ie$ factor and the subindex $0$ denotes this tree level result.

A general parametrization of the vertex fullfiling the C, P and CP symmetries is given by\cite{hagiwara,nieves}:
\begin{equation}\label{effvertex}
\Gamma_{eff}^{\lambda\mu\nu}=g_{1} \left(k_{-}-k_{+}\right)^{\lambda}g^{\mu\nu}+( g_{1}+\kappa+\lambda)\left(q^{\nu}g^{\lambda\mu}-q^{\mu}g^{\nu\lambda}\right)-
\frac{\lambda}{m_{W}^{2}}\left(k_{-}-k_{+}\right)^{\lambda} q^{\mu}q^{\nu}.
\end{equation}
 Comparing the above expressions we identify that for the SM the effective parameters are:
 $g_{1}=1$,  $1+\kappa+\lambda=2$ and $\lambda=0$. In the  static limit, such parameters are related to the electromagnetic multipoles as follows: $\mathcal{Q} = g_1=1$ is the electric charge (in $e$ units), $\mu_W=1+\kappa+\lambda=2$ is the magnetic dipole moment (in $e/2 M_W$ units)  and the electric quadrupole is $X_E =-(\kappa-\lambda)=-1$ (in $e/M_W^{2}$ units). These values are usually taken as a reference for vector mesons. We have also shown that the analog of Burnett and Kroll theorem \cite{bk} for polarized radiative interferences favors the  $|\vec{\mu}|=2$ value \cite{g2}. In Figure \ref{vertex}, we show the generic vertex of a vector particle and the photon.\\
 
The measurement performed, for example, by The DELPHI Collaboration \cite{delphi} at LEP2 considered the  $e^+e^- \to 4 fermions$ process, where the four fermions were associated to a $W^+W^-$ production (Figure \ref{mainmode}), which then decay to pairs of fermions, identified as jets and/or leptons. The observations were found to be consistent with the SM values.

In this proceedings, we describe a similar approach for vector mesons which mainly decay into two pseudoscalars ($P$). Thus, processes of the form $e^{+} e^{-} \to P^+ P^-  P^0 P^0$ can be useful to explore the electromagnetic structure of the vector mesons. We devote most of the work to discuss the general features of the elements relevant for the description of the process and we close with particular cases for light vector mesons, namely for the $\rho$ and $K^*$ mesons.

\section{The elements in the $e^{+} e^{-} \to  P^+ P^-  P^0 P^0$  process}
\begin{figure}
\begin{center}
\begin{minipage}{16pc}
\includegraphics[width=16pc]{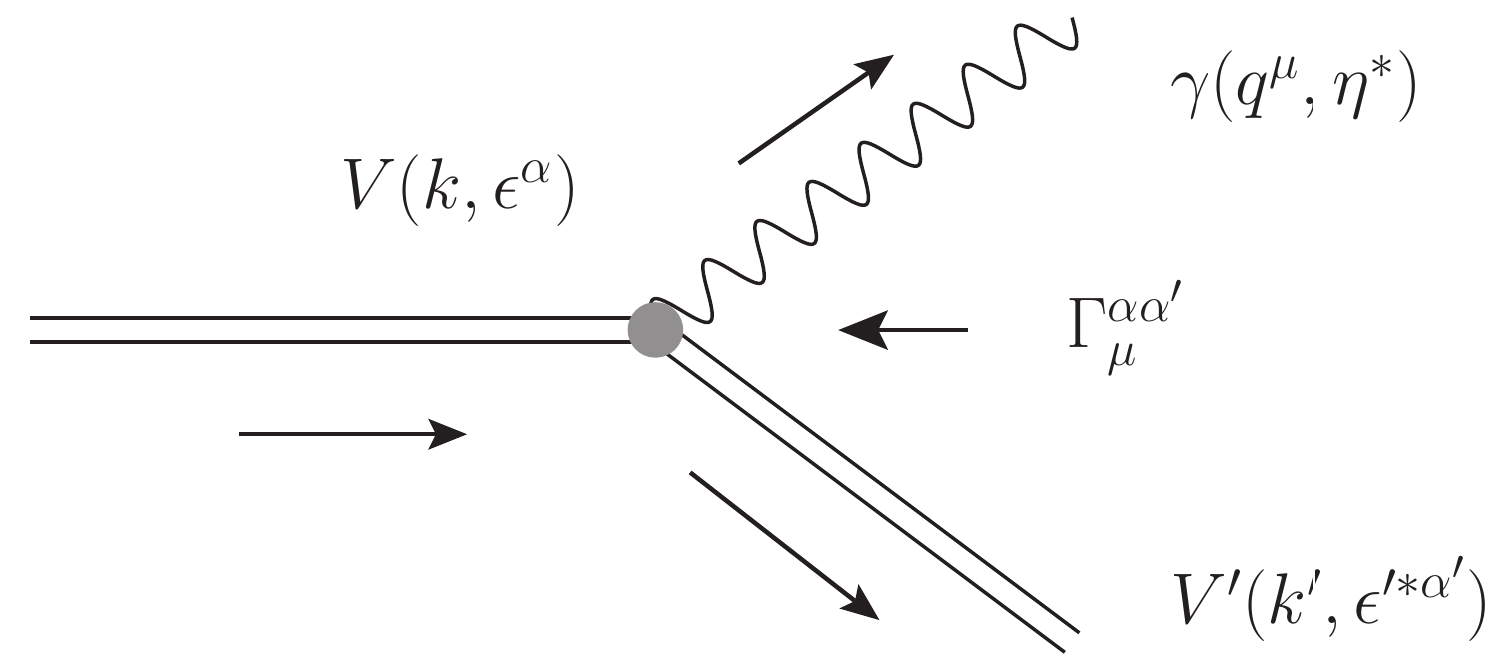}
\caption{\label{vertex} Vector meson electromagnetic vertex, in a generic convention. Within parenthesis are the corresponding momentum and  polarization tensor.}
\end{minipage}
\hspace{2pc}%
\begin{minipage}{16pc}
\includegraphics[width=16pc]{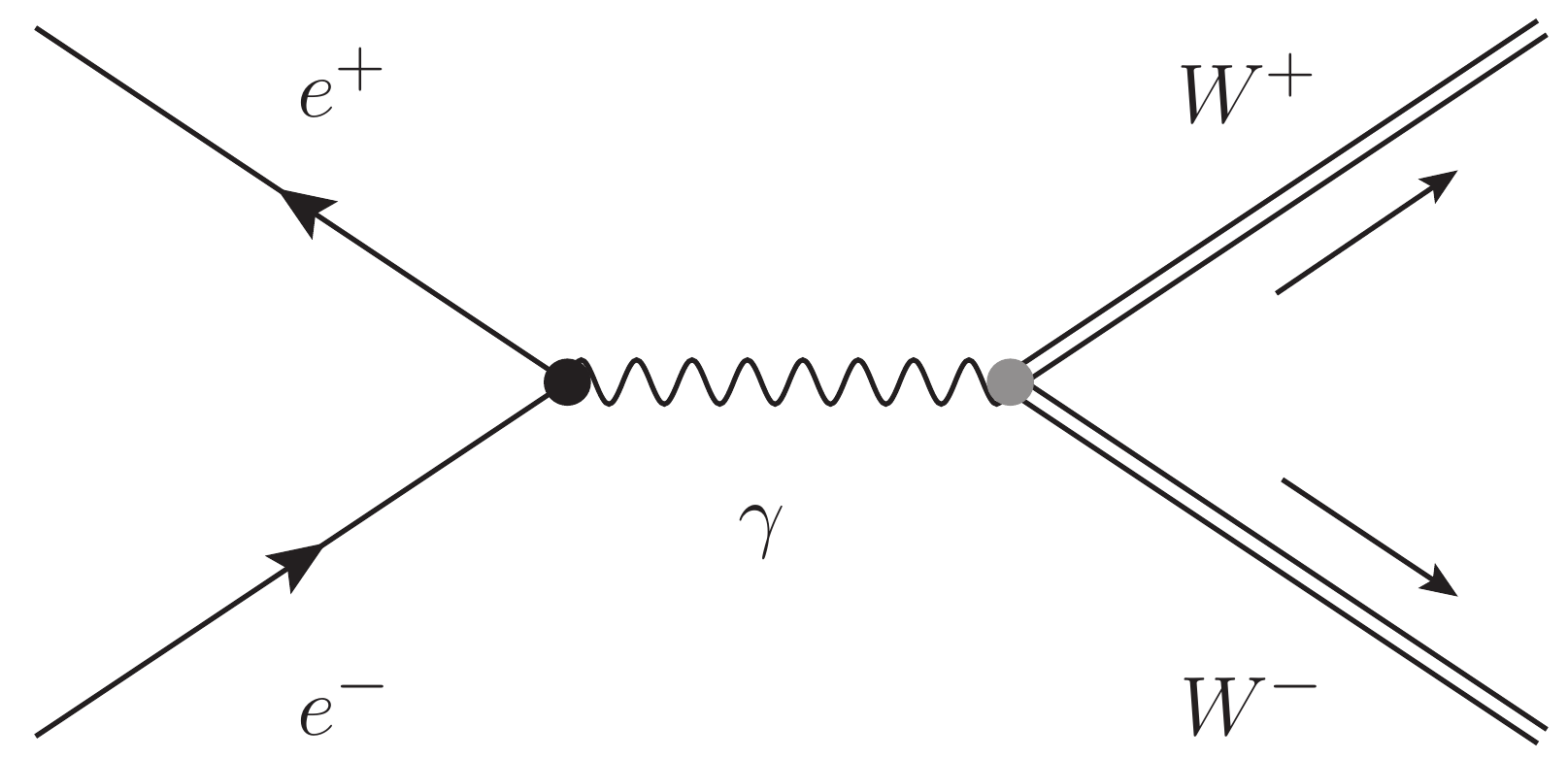}
\caption{\label{mainmode}$W$ gauge boson pair production. The full  $e^+e^- \to 4 fermions$ process has the analog $e^{+} e^{-} \to  P^+ P^-  P^0 P^0$  process for vector mesons.}
\end{minipage} 
\end{center}
\end{figure}

Let us discuss each element relevant for the description of the probability amplitude of this process.  The electron-positron annihilation into a photon is properly accounted by a QED description. 
The description of the hadronic part relevant for our purposes considers that the photon becomes a neutral vector meson, based on the vector meson dominance model (VMD), which then produces two charged vector mesons, each decaying into two pseudoscalar mesons, as depicted in Figure \ref{mainmode}. The following elements are important in order to account for the probability\\

\begin{itemize}
\item The vector meson propagator and the unstable feature
\item The three vector mesons vertex 
\item Bose-Einstein and Charge conjugation symmetries 
\item Gauge invariance of the amplitude
\end{itemize}

Let us address each point:
\subsection{The vector mesons propagator and the unstable feature}
The finite decay width ($\Gamma$) of the $W$ became relevant with the advent of colliders capable of producing this gauge boson on-shell. In order to avoid the propagator pole, several prescriptions were introduced on phenomenological grounds. The so-called complex-mass scheme \cite{complexmass,complexmass2} succeeded in this task while fulfilling the electromagnetic gauge invariance condition.  The fermion-loop scheme \cite{fermionloop,fermionloop2,fermionloop3}  provided a proper quantum field theory ground by accounting for the absorptive contributions of fermions to the electromagnetic vertex and the propagator, linked by the Ward-Takahashi identity.\\ 
These schemas are consistent with each other upon renormalization of the gauge field. Thus, it is common to consider the $W$ propagator as given in the complex-mass scheme:
 \begin{equation}
 D^{\mu\nu}[q^2,W]=i\left(\frac{-g^{\mu\nu}+\frac{q^\mu q^\nu}{M_W-iM_W\Gamma} }{q^2-M_W^2+iM_W\Gamma}\right),
 \label{vectorpropagator}
 \end{equation}
where the width can be taken as either constant or momentum dependent.

In the case of vector mesons, the same prescription can be translated to account for their finite width by noticing that the decay width is driven by the decay into two pseudoscalars. The boson-loop scheme \cite{glctau2pi} considered the absorptive contributions of such particles. Splitting the absorptive correction into transverse ($\mathrm{Im} \Pi^{T}$) and longitudinal ($\mathrm{Im} \Pi^{L}$) ones, the propagator in general becomes \cite{glctau2pi}:

\begin{equation}\label{propagador_s_12}
D^{\mu\nu }[q^2,V]  =\frac{i}{q^{2}-M_V^{2}+i%
	\mathrm{Im}\Pi ^{T} }\left\{ -g^{\mu\nu
}+q^{\mu }q^{\nu }\frac{q^2+i\left( \mathrm{Im}\Pi
^{T} -\mathrm{Im} \Pi^{L} \right) }{%
q^{2}\left( M_V^{2}-i\mathrm{Im}\Pi^{L} \right) }%
\right\},   
\end{equation}%
where $\mathrm{Im} \Pi^{T}$ is directly proportional to the width and $\mathrm{Im} \Pi^{L}$ to the squared mass difference of the bosons in the loop. Notice that for the $\rho$ the particles in the loop are the $\pi^+$ and $\pi^0$, thus the longitudinal contribution is nearly zero. However, for the $K^*$ they correspond to  $K^+\pi^0$ and $K^0\pi^+$  and important corrections from $SU(3)_F$ symmetry breaking are expected \cite{luis}.

\subsection{The three vector mesons vertex}  
The fact that there is no resonant state coupled to the photon producing the $W$ pair allows to directly identify the couplings with the electromagnetic properties of the $W$. However, the electromagnetic structure of the vector meson as a function of the momentum can be accomplished by the inclusion of resonances coupled to the photon. The structure of the triple vector meson vertex is expected to follow the same form as the electromagnetic vertex previously discussed but for a set of different couplings, such that the electric charge form factor is written as:
\begin{equation}
 \left(k_{-}-k_{+}\right)^{\lambda}g^{\mu\nu} \to \sum_{j=V'} \frac{-1}{g_{j}}
\frac{g_{jVV }M_j^2}{q^2-M_j^2+ i M_j \Gamma _j} \left(k_{-}-k_{+}\right)^{\lambda}g^{\mu\nu},
\label{fformarec}
\end{equation}
where $j$ runs over all the possible resonances, and the $g_j$ and $g_{jVV}$  are related to the photon and vector meson couplings of such resonances respectively. 
The condition that at $q^2\to 0$ both expressions must be the same, corresponding to the electric charge of the particle, imposes a relation between the couplings. 

 We have computed the correction to the multipoles of the $W$, $\rho$ and $K^*$ particles, exclusively from the fact they are unstable and found them to be relatively small \cite{david:2010}. Thus, such correction is usually not included.
 
\subsection{Bose-Einstein and Charge conjugation symmetries} 
Now, we get into the properties of the amplitude of the process. We take as a definite case the production of the four pseudoscalars to illustrate the conditions to be fulfilled due to the nature of the particles produced and the electromagnetic origin. This is particularly useful for effective theories. In the case of the four fermions, the SM construction takes care of many of the details.

Let us set our notation as: $ e^{+}(k_1) e^{-}(k_2) \to P^+(p_1) P^0(p_2) P^-(p_3) P^0(p_4)$, in parenthesis are the corresponding 4-momenta. The total amplitude can be written as:
\begin{equation}
\mathcal{M}=\frac{-ie }{(k_1+k_2)^2} l^\mu h_\mu(p_1, p_2 ,p_3, p_4) ,
\end{equation}
where the leptonic current is given by $l^{\mu} \equiv  \bar{v}(k_2) \gamma^\mu u(k_1)$ and  $h_\mu$ represents the four pseudoscalars electromagnetic current. In general, this is not limited to the channel where the pair of intermediate vector mesons is produced. The fact that it involves identical neutral particles imposes the requirement of Bose-Einstein symmetry. That is, the amplitude must be symmetric under the exchange of the identical neutral particles. This is represented by:
\begin{equation}
h_\mu (p_1, p_2 ,p_3, p_4)=h_\mu(p_1, p_4 ,p_3, p_2).
\end{equation}
In addition, the current must be symmetric under charge conjugation ($C$ invariance). In  the case that the neutral particles are their own anti-particles, and the charged ones are the antiparticle of each other, it corresponds to the exchange of them:
\begin{equation}
h_\mu (p_1, p_2 ,p_3, p_4)=- h_\mu (p_3, p_2 ,p_1, p_4).
\end{equation}
Thus, in order to account for such requirements, the total contribution  can be written as the sum of the four possible momenta configurations, represented by a reduced amplitude $\mathcal{M}_{r \mu}$ no longer constrained by such symmetries \cite{ecker}:
\begin{eqnarray}
h_\mu (p_1, p_2 ,p_3, p_4) &=& \mathcal{M}_{r \mu} (p_1, p_2 ,p_3, p_4) + \mathcal{M}_{r \mu} (p_1, p_4 ,p_3, p_2)  \nonumber \\
&-&\mathcal{M}_{r \mu} (p_3, p_2 ,p_1, p_4) - \mathcal{M}_{r \mu} (p_3, p_4 ,p_1, p_2).
\label{ampcanaltot}
\end{eqnarray}

Note that, since the different channels add up to build the total amplitude, we can impose this construction channel by channel. An example of the reduced amplitude is the one involving the intermediate pair of vector mesons:
 
	\begin{equation}\label{Mr_1}
	\mathcal{M}_{r}^{\mu}(p_{1},p_{2},p_{3},p_{4})=(p_{2}-p_{1})_{\lambda}D^{\delta \lambda}[s_{12}, V^{+}]\Gamma_{\gamma \delta \epsilon}D^{\epsilon \nu}[s_{34},V^{-}](p_{4}-p_{3})_{\nu}D^{\mu\gamma}[q^2,V'],
	\end{equation}

where we have defined $s_{ij}\equiv (P_i+P_j)^2$ and the derivative coupling of the vector meson to the pseudoscalars was used.

\subsection{Gauge invariance of the amplitude}
In general, not all the diagrams contributing to the amplitude are independent, gauge invariance condition requires some of them to be related in such a way that the condition is fulfilled. The leptonic current is certainly gauge invariant. Thus, the condition is translated to the hadronic part, which can be stated as follows:

\begin{equation}\label{qh_0}
q^{\mu}h_\mu (p_1, p_2 ,p_3, p_4) =0.
\end{equation}
In a general case, this condition can be broken and the current can be split into two parts, one which satisfies the condition ($GI$) and other which fails to do so ($NGI$):

\begin{equation}\label{qMr_1}
q^{\mu}h_\mu (p_1, p_2 ,p_3, p_4) =q^{\mu}h^{GI}_\mu (p_1, p_2 ,p_3, p_4)+q^{\mu}h^{NGI}_\mu (p_1, p_2 ,p_3, p_4) 0\ne 0,
\end{equation} 
Thus, in order to have a fully gauge invariant amplitude, a counter term must be added to the original amplitude, built out from the $NGI$ contraction:

\begin{equation}
h^{Total}_\mu (p_1, p_2 ,p_3, p_4) =h_\mu (p_1, p_2 ,p_3, p_4)+ h^{CT}_\mu (p_1, p_2 ,p_3, p_4),
\end{equation}
such that now Eqn. (\ref{qh_0}) is fulfilled.

Due to the charge conservation relation of the gauge invariance, contributions coming from charge couplings from different diagrams are related  to this level by gauge invariance. Higher order terms are usually introduced as gauge invariant by themselves. An explicit example of a higher order gauge invariant term is the magnetic dipole moment structure in the electromagnetic vertex Eqn. (\ref{effvertex}). Certainly, these can be worked out at the reduced amplitude level.

\begin{figure}
\begin{center}
\begin{minipage}{16pc}
\includegraphics[width=17pc]{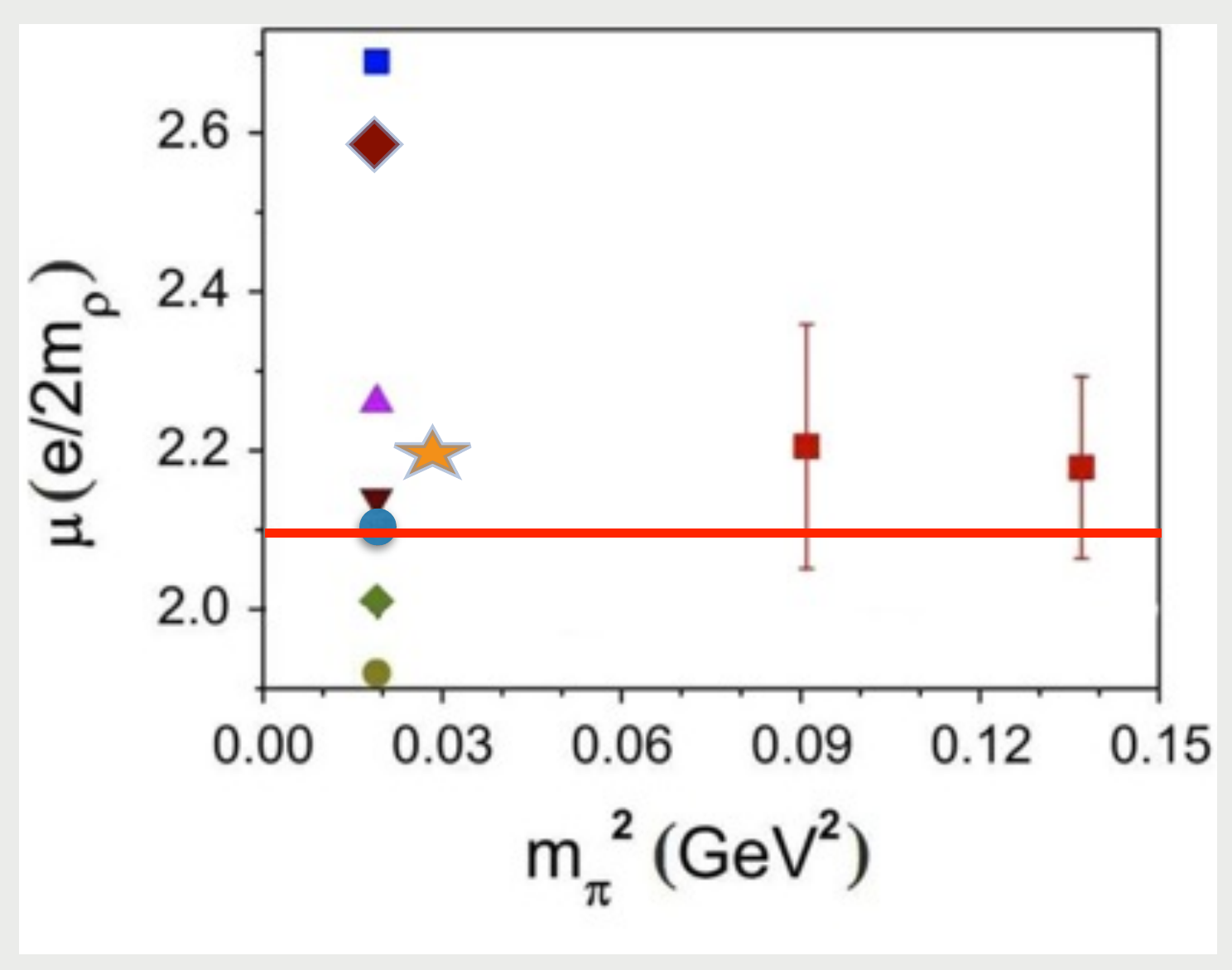}
\caption{\label{rhomdm}$\rho$ Magnetic dipole moment value determined from BaBar data (solid line) compared with the predictions from effective models of QCD and Lattice results at several pion mass values (symbols).}
\end{minipage} 
\hspace{2pc}%
\begin{minipage}{16pc}
\includegraphics[width=15pc]{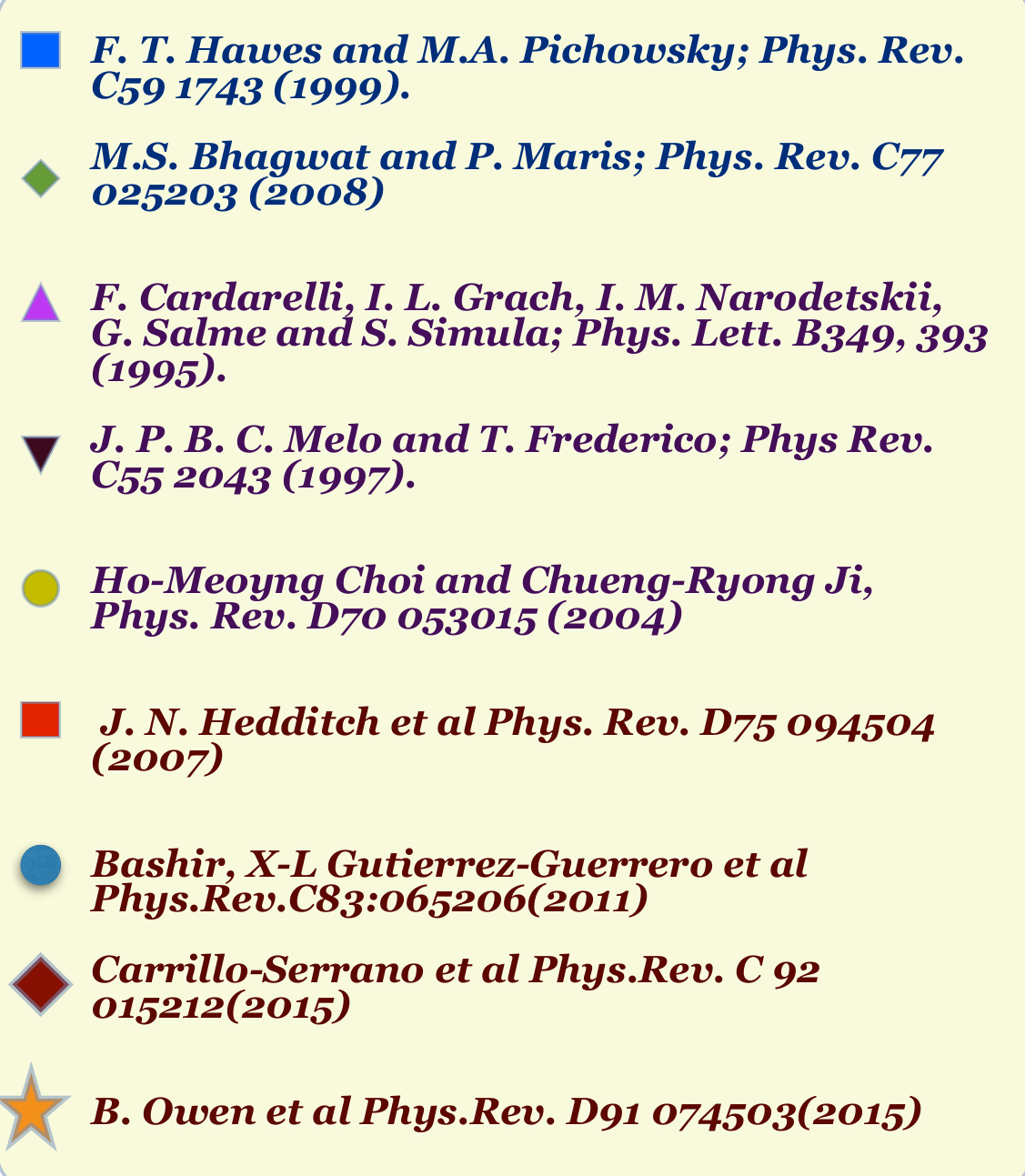}
\caption{\label{refmdm}References for the symbols in Figure \ref{rhomdm}. }
\end{minipage}\hspace{2pc}%
\end{center}
\end{figure}

\section{Light vector mesons}
Experimental information exist on the $e^{+} e^{-} \to \pi^+ \pi^- 2 \pi^0$ process in the low energy regime \cite{snd4pi,ee4pi,ee4pi2,ee4pi3,ee4pi4,ee4pi5,ee4pi6}. The leading channel contribution to the total cross section involves the $\omega-\pi$ mesons intermediate state. Measurements at increasing energies made clear that this process was also sensitive to the channel where the pions are produced in pairs through the $\rho-\rho$ meson resonant states \cite{babar4pi}. This particularity opens the door to explore the $\rho$ electromagnetic vertex and by consequence the corresponding electromagnetic multipoles.
The general procedure discussed above, has been used to determine the magnetic dipole moment of the $\rho$ meson, by studying its effect in the $e^+ e^- \to \pi^+ \pi^- 2 \pi^0$ cross section. The hadronic part was modeled in the VMD approach, which involves several channels. Using preliminary BaBar data  \cite{babar4pi}, while using other observables to fix all the remaining parameters, we have found \cite{mdm,IJMPA} that the best fit  implies a value of $ \mu_\rho=  2.1 \pm 0.5 \: [\frac{e}{2 m_\rho}] $.  The quoted error bar takes into account the uncertainties coming from the couplings of the different channels and model assumptions, dominated  by the lack of information on the $rho«$ properties. In Figure \ref{rhomdm} we compare the central value of this result with the predictions from effective models of QCD and the  Lattice result at several pion mass values \cite{pred1,pred2,pred3,pred4,pred5,pred6,pred7,pred8,pred9}.\\
Another example is the  $e^{+} e^{-} \to K^+ K^- 2 \pi^0$ process, where an intermediate pair of $K^*$ vector mesons are produced. Experimental information on this process still has large uncertainties \cite{kstardata}, but the experimental capabilities are certainly increasing. In this particular case, the $SU(3)_F$ symmetry breaking for the vector meson is large and might have an appreciable effect.

 \section{Conclusions}
 We have described the general procedure to study the $e^{+} e^{-} \to P^+ P^-  P^0 P^0$ in order to gain insight into the properties of vector mesons. Using the $W$ gauge boson analog, as a baseline, we discussed each element relevant for the description of the probability amplitude of this process, paying particular attention to the vector meson electromagnetic vertex and its unstable feature, where not only the finite decay width is important but also isospin symmetry breaking corrections might be explored. An application of the formalism to determine the magnetic dipole moment of the $\rho$ meson has proven the sensitivity of this process to the electromagnetic structure of the vector mesons and open the door to explore similar processes like that involving the $K^*$ meson. 
Finally, we would like to stress the importance of having fine data on these kind of processes to improve our knowledge of the vector mesons.\\ 

\begin {acknowledgments}

We congratulate the Mexican Division of Particles and Fields for its $30^{th}$ annual meeting, and thank the organizers for a very pleasant gathering.
\end {acknowledgments}

\end{document}